\documentstyle[12pt]{article}

\newcommand{\mtx}[2]{\left(\begin{array}{#1}#2\end{array}\right)}

\begin{document}

\begin{center}

\bigskip
{\Large Entangled Chains}\\

\bigskip

William K.~Wootters\\

\bigskip

{\small{\sl

Department of Physics, Williams College, Williamstown, 
MA 01267, USA }}\vspace{3cm}

\end{center}
\subsection*{\centering Abstract}
{Consider an infinite collection of
qubits arranged in a line, such that every pair of nearest neighbors
is entangled: an ``entangled chain.''  
In this paper 
we consider entangled chains with translational
invariance and ask how large one can make the nearest 
neighbor entanglement.  We find that it is possible 
to achieve an entanglement of formation equal to 0.285
ebits between each pair of nearest neighbors, and that 
this is the best one can do under certain assumptions 
about the state of the chain.}

\vfill

PACS numbers: 03.67.-a, 03.65.Bz, 89.70.+c\vfill

\newpage

\section{Introduction: Example of an entangled chain}
Quantum entanglement has been studied for decades, first because of its 
importance in the foundations of quantum mechanics, and more recently for its 
potential technological applications as exemplified by a quantum computer \cite{DiVincenzo}.  The 
new focus has led to a quantitative theory of entanglement \cite{BBPS,Vedral,BDSW} that, among other 
things, allows us to express analytically the degree of entanglement between 
simple systems \cite{HillWootters}.  This development makes it possible to pose new quantitative 
questions about entanglement that could not have been raised before and that 
promise fresh perspectives on this remarkable phenomenon.  In this paper I would
like to raise and partially answer such a question, concerning the extent to 
which a collection of binary quantum objects (qubits) can be linked to each 
other by entanglement.  

Imagine an infinite string of qubits, such as two-level atoms or the spins of 
spin-1/2 particles.  Let us label the locations of the qubits with an integer 
$j$ that runs from negative infinity to positive infinity.  I wish
to consider special states of the string, satisfying the following two 
conditions: (i) each qubit is entangled with its nearest neighbors; (ii) the state is invariant under all translations, that is, under transformations that shift each qubit from its original position $j$ to position $j+n$ for some integer $n$.   Let us call a string of qubits satisfying the first condition an entangled chain, and if it also satisfies the second condition, a uniform entangled chain.  Note that each qubit need not be entangled with any qubits other than its two nearest neighbors.  In this respect an entangled chain is like an ordinary chain, whose links are directly connected only to two neighboring links.  By virtue of the translational invariance, the degree of entanglement between nearest neighbors in a uniform entangled chain must be constant throughout the chain.  The main question I wish to pose is this: How large can the nearest-neighbor entanglement be in a uniform entangled chain? 

This problem belongs to a more general line of inquiry about how entanglement
can be shared among more than two objects.  Some work on this subject has been done in the context of the cloning of entanglement [\textbf{6--11}],
where one finds limits on the extent to which entanglement can be copied.
In a different setting not particularly involving cloning, one finds an inequality bounding the amount of entanglement that a single qubit can have with each of two other qubits \cite{Coffman}.  One can imagine more general ``laws of entanglement sharing'' that apply to a broad range of configurations of quantum objects.  The present work provides further data that might be used to discover and formulate such laws.    
The specific problem addressed in this paper could also prove relevant for analyzing models of quantum computers in 
which qubits are arranged along a line, as in an ion trap \cite{Zoller}.  The infinite chain 
can be thought of as an idealization of such a computer.  Moreover, the analysis of our question turns out to be interesting in its own right, being related, as we will see, to a familiar problem in many-body physics.  

To make the question precise we need a measure of entanglement between two 
qubits.  We will use a reasonably simple and well-justified measure called the 
``concurrence,'' which is defined as follows \cite{HillWootters}.  

Consider first the case of pure states.  A general
pure state of two qubits can be written as
\begin{equation}
|\psi\rangle = \alpha |00\rangle + \beta |01\rangle + \gamma |10\rangle + \delta |11\rangle.
\end{equation}
One can verify that such a state is factorizable into single-qubit states---that is, it is unentangled---if and only if $\alpha \delta = \beta \gamma$.  The quantity \hbox{$C = 2|\alpha\delta-\beta\gamma|$}, which ranges from 0 to 1, is thus a plausible measure of the degree of 
entanglement.  We take this expression as the
definition of concurrence for a pure state of two qubits.  For mixed states,
we define the concurrence to be the greatest convex function on the set of 
density matrices that gives the correct values for pure states \cite{Uhlmann}.

Though this statement defines concurrence, it does not tell us how to 
compute it for mixed states.  
Remarkably, there exists an explicit formula for the concurrence of an 
arbitrary mixed state of two qubits \cite{HillWootters}:  Let $\rho$ be the density matrix 
of the mixed state, which we imagine expressed in the standard basis 
$\{|00\rangle, |01\rangle, |10\rangle, |11\rangle\}$. Let $\tilde \rho$, the 
``spin-flipped'' density matrix, be \hbox{$(\sigma_y\otimes\sigma_y)\rho^* (\sigma_y\otimes \sigma_y)$}, 
where the asterisk denotes complex conjugation in the standard basis and 
$\sigma_y$ is the matrix $\mtx{cc}{0&{-i}\\i&0}$.  Finally, let $\lambda_1,
\lambda_2, \lambda_3, \lambda_4$ be the square roots of the eigenvalues of $\rho
\tilde \rho$ in descending order---one can show that these eigenvalues are all 
real and non-negative.  Then the concurrence of $\rho$ is given by 
the formula
\begin{equation}
C(\rho) = {\rm max}\{\lambda_1-\lambda_2-\lambda_3-\lambda_4,0\}.
\end{equation}
The best justification for using concurrence as a measure of entanglement comes from a theorem \cite{HillWootters} showing that concurrence is a monotonically increasing function of the ``entanglement of formation,'' which quantifies the 
non-local resources needed to create the given state \cite{BDSW}.\footnote{One can define the entanglement of formation as follows.  Let $\rho$ be a mixed state of a pair of quantum objects, to be shared between two separated observers who can communicate with each other only via classical signals.  The entanglement of formation of $\rho$ is the asymptotic number of singlet states the observers need, per pair, in order to create a large number of pairs in {\em pure} states whose average density 
matrix is $\rho$.  (This is conceptually different from the {\em regularized} entanglement of formation, which measures the cost of creating many copies of the {\em mixed} state $\rho$ \cite{Terhal}. 
However, 
it is conceivable that the two quantities are identical.)  Entanglement of formation is conventionally measured in ``ebits,'' and for a pair of binary quantum objects it takes values ranging from 0 to 1 ebit.}  As mentioned above, the 
values of $C$ 
range from zero to one: an unentangled state has
$C=0$, and a completely entangled state such as the singlet state 
$\frac{1}{\sqrt{2}}(|01\rangle - |10\rangle)$ 
has $C=1$.  Our problem is to
find the greatest possible nearest-neighbor concurrence of a uniform entangled chain.  
At the end of the calculation we can easily re-express our results in terms of 
entanglement of formation. 

Another issue that needs to be addressed in formulating our question is the 
meaning of the word ``state'' as applied to an infinite string of qubits; in 
particular we need to discuss how such a state is to be normalized.  Formally, 
we can define a state of our system as follows.  A state $w$ of the infinite 
string is a function that assigns to every finite set $S$ of integers a 
normalized ({\em i.e.}, trace one) density matrix $w(S)$, which we interpret 
to be the density matrix of the qubits specified by the set $S$; moreover the 
function $w$ must be such that if $S_2$ is a subset of $S_1$, then $w(S_2)$ 
is obtained from $w(S_1)$ by tracing over the qubits whose labels are not in
$S_2$.  This formal definition is perfectly sensible but somewhat bulky in 
practice.  In what follows we will usually specify states of the string more 
informally when it is clear from the informal specification how to 
generate the density matrix of any finite subset of the string.  We will 
also usually use the symbol $\rho$ instead of $w(S)$ to denote the density matrix of a pair of nearest neighbors.

It is not immediately obvious that there exists even a single example of an 
entangled chain.  Note, for example, that the limit of a Schr\"{o}dinger cat 
state---an equal superposition of an infinite string of zeros with an infinite 
string of ones---is not an entangled chain.  In the cat state, the reduced density matrix of a pair of 
neighboring qubits is an incoherent mixture of $|00\rangle$ and $|11\rangle$, 
which exhibits a classical correlation but no entanglement.  (Note, by the way, 
that our informal statement ``an equal {\em superposition} of an infinite string
of zeros with an infinite string of ones,'' specifies exactly the same state as 
if we had taken an incoherent {\em mixture} of these two infinite strings: no 
finite set of qubits contains information about the phase of the superposition.)

We can, however, construct a simple example of an entangled chain in the following way.  
Let $w_0$ be the state such that for each {\em even} integer $j$, the qubits at 
sites $j$ and $j+1$ are entangled with each other in a singlet state.  We can 
write this state informally as\footnote{Alternatively, we can characterize the 
state $w_0$ according to our formal definition by specifying the density matrix 
of each finite collection of qubits:  Let $S$ define such a collection.  Then 
for each even integer $j$ such that both $j$ and $j+1$ are in $S$, the 
corresponding pair of qubits is in the singlet state; all other qubits ({\em 
i.e.}, the unpaired ones) are in the completely mixed state 
$\mtx{cc}{\frac{1}{2}&0\\0&\frac{1}{2}}$, and the full density matrix $w(S)$ is 
obtained by taking the tensor product of the pair states and single-qubit 
states.}
\begin{equation}
\cdots \otimes 
\left(\frac{|0\rangle_{-2}|1\rangle_{-1}-|1\rangle_{-2}|0\rangle_{-1}}{\sqrt{2}}\right)
\otimes \left(\frac{|0\rangle_{0}|1\rangle_{1}-|1\rangle_{0}|0\rangle_{1}}{\sqrt{2}}\right)
\otimes \left(\frac{|0\rangle_{2}|1\rangle_{3}-|1\rangle_{2}|0\rangle_{3}}{\sqrt{2}}\right)
\otimes \cdots .
\end{equation}
The state $w_0$ is not an entangled chain because the qubits are not entangled with both of their nearest 
neighbors: qubits at even-numbered locations are not entangled with their 
neighbors on the left.  However, if we let $w_1$ be the state obtained by 
translating $w_0$ one unit to the left (or to the right---the result is the 
same), and let $w$ be an equal mixture of $w_0$ and $w_1$---that is, 
$w=(w_0+w_1)/2$---then $w$ is a uniform entangled chain, as we now show.

That $w$ is translationally invariant follows from the fact that both $w_0$ and 
$w_1$ are invariant under {\em even} displacements and that they transform into 
each other under odd displacements.  Thus we need only show that 
neighboring states are entangled.  For definiteness let us consider the qubits 
in locations $j=1$ and $j=2$.  In the state $w_0$, the density matrix for these 
two qubits is
\begin{equation}
\rho^{(0)} = \mtx{cccc}{\frac{1}{4}&0&0&0\\0&\frac{1}{4}&0&0\\0&0& 
\frac{1}{4}&0\\0&0&0&\frac{1}{4}},
\end{equation}
that is, the completely mixed state.  (The two qubits are from distinct singlet 
pairs.)  The density matrix of the same two qubits in the state $w_1$ is 
\begin{equation}
\rho^{(1)} = \mtx{cccc}{0&0&0&0\\0&\frac{1}{2}&-\frac{1}{2}&0\\
0&-\frac{1}{2}&\frac{1}{2}&0\\0&0&0&0},
\end{equation}
that is, the singlet state.  In the state $w$, the qubits are in an equal 
mixture of these two density matrices, which is
\begin{equation}
\rho = (\rho^{(0)}+\rho^{(1)})/2 = 
\mtx{cccc}{\frac{1}{8}&0&0&0\\
0&\frac{3}{8}&-\frac{1}{4}&0\\
0&-\frac{1}{4}&\frac{3}{8}&0\\
0&0&0&\frac{1}{8}}.   \label{bike}
\end{equation}
It is easy to compute the concurrence of this density matrix, because 
$\tilde{\rho}$ is the same as $\rho$ itself.  The values 
$\lambda_i$ in this case are the eigenvalues of $\rho$, which are $\frac{5}{8}, 
\frac{1}{8}, \frac{1}{8}, \frac{1}{8}$.  The concurrence is therefore 
$C=\frac{5}{8}- \frac{1}{8} -\frac{1}{8} - \frac{1}{8}=\frac{1}{4}$.  This same 
value of the concurrence applies to any other pair of neighboring qubits in the 
string because of the translational invariance.  The fact that the concurrence 
is non-zero implies that neighboring qubits are entangled, so that the state $w$
is indeed an entangled chain.  For uniform entangled chains, we will call the common value of $C$ for 
neighboring qubits the concurrence of the chain.  Thus in the above example 
the concurrence of the chain is $\frac{1}{4}$.    

As we will see, it is possible to find uniform entangled chains with greater 
concurrence.  Let $C_{\rm max}$ be the least upper bound on the concurrences of all 
uniform entangled chains.  We would like to find this number.  We know that $C_{\rm max}$ is
no larger than 1, since concurrence never exceeds 1.  In fact we can quickly get
a somewhat better upper bound, using the following fact: when a qubit is entangled 
with each of two other qubits, the sum of the squares of the two concurrences is
less than or equal to one \cite{Coffman}.  In a uniform entangled chain, each qubit must be equally 
entangled with its two nearest neighbors; so the concurrence with each of them 
cannot exceed $1/\sqrt{2}$.  Thus, so far what we know about $C_{\rm max}$ is this:
\begin{equation}
1/4 \leq C_{\rm max} \leq 1/\sqrt{2}.
\end{equation}
This is still a wide range.  Most of the rest of this paper is devoted to 
getting a better fix on $C_{\rm max}$ by explicitly constructing entangled chains.

\section{Building chains out of blocks}
Using the above example as a model, we will use the following construction to 
generate other uniform entangled chains.  (1) Break the string into blocks of $n$ qubits, and define a state $w_0$ in which each block is in the same $n$-qubit 
state $|\xi\rangle$; that is, $w_0$ is a tensor product of an infinite number of
copies of $|\xi\rangle$.  (In the above example $n$ had the value 2 and 
$|\xi\rangle$ was the singlet state.)  (2) Define $w_k$, $k=1,\ldots,n-1$, to 
be the state obtained by shifting $w_0$ to the left by $k$ units.  (3) Let the
final state $w$ be the average $(w_0+\cdots +w_{n-1})/n$.  A state generated in 
this way will automatically be translationally invariant.  In order that the 
chain have a large concurrence, we will need to choose the state $|\xi\rangle$ 
carefully.  Finding 
an optimal $|\xi\rangle$ and proving that it is optimal may turn out to be a difficult problem.  In this paper I 
will choose $|\xi\rangle$ according to a strategy that makes sense and may well 
be optimal but is not proven to be so. 

In the final state $w$, each pair of neighboring qubits has the same density 
matrix because of the translational invariance.  Our basic strategy for choosing $|\xi\rangle$, described 
below, is designed to give this neighboring-pair density matrix the following 
form:
\begin{equation}
\rho = \mtx{cccc}{\rho_{11}&0&0&0\\
0&\rho_{22}&\rho_{23}&0\\
0&{\rho}^*_{23}&\rho_{33}&0\\  \label{form}
0&0&0&0}.
\end{equation}
(The ordering of the four basis
states is the one given above: $|00\rangle, |01\rangle, |10\rangle, 
|11\rangle$.)  One can show that the concurrence of such a density matrix is 
simply 
\begin{equation}
C = 2\big|\rho_{23}\big|.  \label{simpleconc}
\end{equation}
Besides making the concurrence easy to compute, the form (\ref{form}) seems a 
reasonable goal because it picks out a specific kind of entanglement, namely, 
a coherent superposition of $|01\rangle$ and $|10\rangle$, and limits the ways in which this entanglement can be
contaminated or diluted by being mixed with other states.  In particular, the 
form (\ref{form}) does not allow contamination by an orthogonal entangled 
state of the form $\alpha|00\rangle + \beta|11\rangle$---orthogonal entangled states when mixed together 
tend to cancel each other's entanglement---or by the combination of the two 
unentangled states $|00\rangle$ and $|11\rangle$.  If the component $\rho_{44}$ 
were not equal to zero and the form were otherwise unchanged, the concurrence 
would be $C= {\rm max} \{2(|\rho_{23}|- \sqrt{\rho_{11}\rho_{44}}),0\}$; so it is good to make 
either $\rho_{11}$ or $\rho_{44}$ equal to zero if this can be done without 
significantly reducing $\rho_{23}$.  We have chosen to make $\rho_{44}$ equal to
zero.

As it happens, one can guarantee the form (\ref{form}) for the density matrix of
neighboring qubits by imposing the following three conditions on the $n$-qubit 
state $|\xi\rangle$:  (i) $|\xi\rangle$ is an eigenstate of the operator that 
counts the number of qubits in the state $|1\rangle$.  That is, each basis 
state represented in $|\xi\rangle$ must have the same number $p$ of qubits in 
the state $|1\rangle$.  (ii) $|\xi\rangle$ has no component in which two 
neighboring qubits are both in the state $|1\rangle$.  (iii) The $n$th qubit is 
in the state $|0\rangle$.  (This last condition effectively extends
condition (ii) to the boundary between successive blocks.)
Condition (i) guarantees that the density matrix 
$\rho$ for a pair of nearest neighbors is block diagonal, each block 
corresponding to a fixed number of 1's in the pair.  That is, there are two 
single-element blocks corresponding to $|00\rangle$ and $|11\rangle$, and a 2x2 
block corresponding to $|01\rangle$ and $|10\rangle$.  Conditions (ii) and 
(iii) guarantee that $\rho_{44}$ is zero.  The conditions thus give us the form 
(\ref{form}).  We impose these three conditions because they 
seem likely to give the best results; we do not prove that they 
are optimal.  

To illustrate the three conditions and how they can be used, let us consider 
in detail the case where the block size $n$ is 5 and the number $p$ of 1's in 
each block is 2.  (Our strategy does not specify the value of either $n$ or $p$;
these values will ultimately have to be determined by explicit maximization.)  In this 
case, the only basis states our conditions allow in the construction of $|\xi\rangle$ are 
$|10100\rangle$, $|10010\rangle$, and $|01010\rangle$.  Any other basis state 
either would have a different number of 1's or would violate one of conditions 
(ii) and (iii).  Thus we write 
\begin{equation}
|\xi\rangle = a_{13}|10100\rangle + a_{14}|10010\rangle + a_{24}|01010\rangle.
\label{xi}
\end{equation}
The subscripts in $a_{ij}$ indicate
which qubits are in the state $|1\rangle$.  The state $w$ of the infinite string
is derived from $|\xi\rangle$ as described above.  We now want to use 
Eq.~(\ref{xi}) to write the density matrix $\rho$ of a pair of nearest neighbors
when the infinite string is in the state $w$.  For definiteness let us take the 
two qubits of interest to be in locations $j=1$ and $j=2$, and let us take the 
5-qubit blocks in the state $w_0$ to be given by $j=1,\ldots,5$, 
$j=6,\ldots,10$, and so on.  Our final density matrix $\rho$ will be an equal 
mixture of five density matrices, corresponding to the five different 
displacements of $w_0$ (including the null displacement).  

For $w_0$ itself, the qubits at $j=1$ and $j=2$ are the first two qubits of 
$|\xi\rangle$.  The density matrix for these two qubits, obtained by tracing out 
the other three qubits of the block, is
\begin{equation}
\rho^{(0)} = \mtx{cccc}{0&0&0&0\\
0&|a_{24}|^2&{a}_{14}^*a_{24}&0\\
0&a_{14}{a}_{24}^*&|a_{13}|^2+|a_{14}|^2&0\\
0&0&0&0}.
\end{equation}
For $w_1$, the qubits at $j=1$ and $j=2$ are now the second and third qubits of 
the block, since the block has been shifted to the left.  Thus we trace over the
first, fourth, and fifth qubits to obtain
\begin{equation}
\rho^{(1)} = \mtx{cccc}{|a_{14}|^2&0&0&0\\
0&|a_{13}|^2&0&0\\
0&0&|a_{24}|^2&0\\
0&0&0&0}.
\end{equation}
In a similar way one can find $\rho^{(2)}$ and $\rho^{(3)}$:
$$
\rho^{(2)}=\mtx{cccc}{0&0&0&0\\
0&|a_{14}|^2+|a_{24}|^2&{a}_{13}^*a_{14}&0\\  
0&a_{13}{a}_{14}^*&|a_{13}|^2&0\\
0&0&0&0};
\rho^{(3)}=\mtx{cccc}{|a_{13}|^2&0&0&0\\
0&0&0&0\\
0&0&|a_{14}|^2+|a_{24}|^2&0\\
0&0&0&0}.
$$
The density matrix corresponding to $w_4$ is different in that the two relevant 
qubits now come from different blocks: the qubit at $j=1$ is the last qubit of 
one block and the qubit at $j=2$ is the first qubit of the next block.  The 
corresponding density matrix is thus the tensor product of two single-qubit 
states:
$$
\rho^{(4)}=\mtx{cc}{1&0\\0&0} \otimes
\mtx{cc}{|a_{24}|^2&0\\         
0&|a_{13}|^2+|a_{14}|^2}
=\mtx{cccc}{|a_{24}|^2&0&0&0\\
0&|a_{13}|^2+|a_{14}|^2&0&0\\
0&0&0&0\\
0&0&0&0}.
$$
To get the neighboring-pair density matrix corresponding to our final state $w$,
we average the above five density matrices, with the following simple result:
\begin{equation}
\rho=\frac{1}{5}\mtx{cccc}{1&0&0&0\\
0&2&x&0\\
0&{x}^*&2&0\\
0&0&0&0},
\end{equation}
where 
\begin{equation}
x = {a}_{13}^*a_{14}+{a}_{14}^*a_{24}.  \label{examp}
\end{equation}
According to Eq.~(\ref{simpleconc}), the concurrence of the pair is
\begin{equation}
C = \frac{2}{5}\big|{a}_{13}^*a_{14}+{a}_{14}^*a_{24}\big|. 
\label{absoluteC}
\end{equation}

Continuing with this example---$n=5$ and $p=2$---let us find out what values we 
should choose for $a_{13}$, $a_{14}$, and $a_{24}$ in order to maximize $C$.  
First, it is clear that we cannot go wrong by taking each $a_{ij}$ to be real and non-negative---any complex phases could only reduce the absolute value in Eq.~(\ref{absoluteC})---so let us restrict our attention to such values.  To
take into account the normalization condition, we use a Lagrange multiplier 
$\gamma /2$ and extremize the quantity
\begin{equation}
a_{13}a_{14}+a_{14}a_{24}-(\gamma /2)(a_{13}^2+a_{14}^2+a_{24}^2).
\end{equation}
Differentiating, we arrive at three linear equations expressed by the matrix 
equation
\begin{equation}
\mtx{ccc}{0&1&0\\
1&0&1\\
0&1&0}
\mtx{c}{a_{13} \\a_{14}\\a_{24}}
= \gamma \mtx{c}{a_{13}\\a_{14}\\a_{24}}.           \label{3x3}
\end{equation}
Of the three eigenvalues, only one allows an eigenvector with non-negative 
components, namely, $\gamma = \sqrt{2}$.  The normalized eigenvector is
\begin{equation}
\mtx{c}{a_{13}\smallskip \\a_{14}\smallskip \\a_{24}}=\mtx{c}{\frac{1}{2}\smallskip \\ \frac{1}{\sqrt{2}}\smallskip \\ \frac{1}{2}},
\end{equation}
which gives $C=\sqrt{2}/5=0.283$.  This is greater than the value 0.25 that we 
obtained in our earlier example.  

Before generalizing this calculation to arbitrary values of $n$ and $p$, we 
adopt some terminology that will simplify the discussion.  Let us think of the 
qubits as ``sites,'' and let us call the two states of each qubit ``occupied'' 
($|1\rangle$) and ``unoccupied'' ($|0\rangle$).  The states $|\xi\rangle$ that 
we are considering have a fixed number $p$ of occupied sites in a string of $n$ 
sites; so we can regard the system as a collection of $p$ ``particles'' in a 
one-dimensional lattice of length $n$.  Condition (ii) requires that two 
particles never be in adjacent sites; it is as if each particle is an extended 
object, taking up two lattice sites, and two particles cannot overlap.  Thus the number of particles is limited by the inequality $2p \le n$.

\section{Generalization to blocks of arbitrary size}
We now turn to the calculation of the optimal concurrence for general $n$ and 
$p$ assuming our conditions are satisfied.  It will turn out that this calculation can be done exactly.  

For any values of $n$ and $p$, the most general form of $|\xi\rangle$ consistent
with condition (i) is
\begin{equation}
|\xi\rangle =\sum_{j_1<\cdots <j_p} a_{j_1,\ldots,j_p}|j_1,\ldots,j_p \rangle, 
\label{summ}
\end{equation}
where $|j_1,\ldots,j_p \rangle$ is the state of $n$ sites $j=1,\dots,n$ in which
sites $j_1,\ldots,j_p$ are occupied and the rest are unoccupied.  Because of 
conditions (ii) and (iii), $a_{j_1,\ldots,j_p}$ must be zero if two of the 
indices differ by 1 or if $j_p$ has the value $n$.  The coefficients in 
Eq.~(\ref{summ}) satisfy the normalization condition
\begin{equation}
\sum_{j_1<\cdots <j_p} | a_{j_1,\ldots,j_p}| ^2 = 1.
\end{equation}
Going through the same steps as in the above example, we find that in the state 
$w$ the density matrix of any pair of neighboring sites is
\begin{equation}
\rho = \frac{1}{n}\mtx{cccc}{n-2p&0&0&0\\
0&p&y&0\\
0&{y}^*&p&0\\
0&0&0&0},
\end{equation}
where 
\begin{equation}
y = \sum_{q=1}^{p} \sum_{j_1< \cdots <j_p} \sum_{j_1^\prime < \cdots <j_p^\prime} \bigg[ {a}^*_{j_1,\ldots,j_p}
a_{j_1^\prime ,\ldots, j_p^\prime }
\delta_{j_q^\prime ,j_q+1} \prod_{r\ne q} \delta_{j_r^{\prime},j_r}\bigg].
\end{equation}
Here $\delta$ is the Kronecker delta, and we define $a_{j_1,\ldots,j_p}$ to 
be zero if any two of the indices are equal.  In words, $y$ is constructed
as follows: Let two coefficients $a_{j_1,\ldots,j_p}$ and $a_{j_1^\prime 
,\ldots, j_p^\prime }$ be called adjacent if they differ in only one index and 
if the difference in that index is exactly one; then $y$ is the sum of all 
products of adjacent pairs of coefficients, the coefficient with the smaller value of the special index being complex conjugated in each case.  In the above example there were 
only two such products, ${a}_{13}^*a_{14}$ and ${a}_{14}^*a_{24}$; hence the form of 
Eq.~(\ref{examp}).

As before, the concurrence of the chain is equal to $2|\rho_{23}|$; that is, 
$C=(2/n)|y|$.  We want to maximize the concurrence over all possible values of the
coefficients that are consistent with conditions (ii) and (iii).  These 
conditions are somewhat awkward to enforce directly: one has to make sure
that certain of the coefficients $a_{j_1,\ldots,j_p}$ are zero.  However, this problem is easily circumvented by defining a new set of indices.    Let $k_1 = j_1$, $k_2 = j_2 - 1$, $k_3 = j_3 - 2$, and so
on up to $k_p = j_p - (p-1)$, and let $b_{k_1,\ldots,k_p} = a_{j_1,\ldots,j_p}$.
The constraints on the new indices $k_r$ are simply that $0 < k_1 < k_2 < \cdots
< k_p < n^{\prime}$, where $n^{\prime}=n-(p-1)$.  Finally, in place of $|\xi\rangle$, define a new vector
$|\zeta\rangle$: 
\begin{equation}
|\zeta\rangle = \sum_{k_1<\cdots <k_p} b_{k_1,\ldots,k_p}|k_1,\ldots,k_p 
\rangle,
\end{equation}
where $|k_1,\ldots,k_p \rangle$ is the state of a lattice of length $n^\prime-1$ 
in which the sites $k_1,\ldots,k_p$ are occupied.  In effect we have removed 
from the lattice the site lying to the right of each occupied site.  Note that our earlier inequality $2p \le n$ becomes, in terms of $n^\prime$, simply $p \le n^\prime -1$, which reflects the fact that the new lattice has only $n^\prime -1$ sites.  The concurrence is still given by $C=(2/n)|y|$, where 
\begin{equation}
y = \sum_{q=1}^{p} \sum_{k_1< \cdots <k_p} \sum_{k_1^\prime < \cdots <k_p^\prime} \bigg[ {b}^*_{k_1,\ldots,k_p}  \label{yequation}
b_{k_1^\prime ,\ldots, k_p^\prime }
\delta_{k_q^\prime ,k_q+1} \prod_{r\ne q} \delta_{k_r^{\prime},k_r}\bigg].
\end{equation}

We can express $y$ more simply by introducing creation and annihilation operators for each site.  We associate with site $k$ the operators 
\begin{equation}
c_k = \mtx{cc}{0&1\\0&0}\,\, {\rm and}\,\,\,  c^\dag_k = \mtx{cc}{0&0\\1&0},
\label{cdef}
\end{equation}
which are represented here in the basis $\{|0\rangle,|1\rangle\}$.  In terms of these operators, we can write $y$ as
\begin{equation}
y = \langle \zeta | \sum_{k=1}^{n^\prime -2} c^\dag_k c_{k+1}|\zeta\rangle.
\label{yy}
\end{equation}  

Our problem is beginning to resemble the nearest-neighbor tight-binding model for electrons in a one-dimensional lattice.  The Hamiltonian for the latter problem---assuming that the spins of the electrons are all in the same state and can therefore be ignored---can be written as\footnote{In Eq.~(\ref{HH}) 
the operators
$c$ and $c^\dag$ are fermionic, whereas those
defined in Eq.~(\ref{cdef}) are not, because they do not
anticommute when they are associated with different sites.  
We could, however, 
use our $c$'s to define genuinely fermionic operators 
in terms of which
the extremization problem has exactly the same form \cite{Lieb}.}
\begin{equation}
H = - \sum_{k=1}^{n^\prime -2}(c_k^\dag c_{k+1} + c_{k+1}^\dag c_k),
\label{HH}
\end{equation}
where we have taken the lattice length to be the same as in our problem, namely, $n^\prime -1$.  From Eqs. (\ref{yy}) and (\ref{HH}) we see that $\langle \zeta |H|\zeta\rangle = -2\,{\rm Re}(y)$. This expectation value is not quite what we need for the concurrence: the concurrence is proportional to the absolute value of $y$, not its real part.  However, as in our earlier example, for the purpose of maximizing $C$ there is no advantage in straying from real, non-negative values of $b_{k_1,\ldots,k_p}$.  If we restrict our attention to such values, then the absolute value of $y$ is the same as its real part, and we can write the concurrence as 
\begin{equation}
C = -\frac{1}{n} \langle \zeta | H |\zeta\rangle. \label{forpositive}
\end{equation}
Thus, maximizing the concurrence amounts to minimizing the expectation value of $H$, that is, finding the ground state energy of the tight-binding model, as long as the ground state involves only real and non-negative values of $b_{k_1,\ldots,k_p}$.  

The one-dimensional tight-binding model is in fact easy to 
solve \cite{Lieb,many}.  Its ground state is the discrete analogue 
of the ground state of a collection of $p$ 
non-interacting fermions in a one-dimensional box. In our case the ``walls'' of the box, where the wavefunction goes to zero, are at $k=0$ and $k=n^\prime$,  and the ground state $|\zeta_0\rangle$ is given by the following antisymmetrized product of sine waves:
\begin{equation}
b_{k_1,\ldots,k_p} \propto {\mathcal A}\Big[\sin\Big( \frac{\pi k_p}{n^\prime}\Big)
\sin\Big( \frac{2\pi k_{p-1}}{n^\prime}\Big)\cdots
\sin\Big( \frac{p\pi k_1}{n^\prime}\Big)\Big].  \label{ground}
\end{equation}
Here ${\mathcal A}$ indicates the operation of antisymmetrizing over the indices
$k_1,\ldots,k_p$.  In the range of values we are allowing for these
indices, that is, $0 < k_1 < k_2 < \cdots
< k_p < n^{\prime}$, the coefficients $b_{k_1,\ldots,k_p}$ are indeed non-negative,
so that Eq.~(\ref{forpositive}) is valid.

The ground state energy, from which we can find the concurrence, is simply the sum of the first $p$ single-particle eigenvalues of $H$.  There are exactly
$n^\prime -1$ such eigenvalues, one for each dimension of the single-particle
subspace; they are given by 
\begin{equation}
E_m = -2\cos\Big(\frac{m\pi}{n^\prime}\Big), \,\, m=1,\ldots,n^\prime -1.
\end{equation}
Thus the concurrence is
\begin{equation}
C = -\frac{1}{n}\langle\zeta_0|H|\zeta_0\rangle = \frac{2}{n}\sum_{m=1}^p
\cos\Big(\frac{m\pi}{n^\prime}\Big).
\end{equation}
Doing the sum is straightforward, with the following result: 
\begin{equation}
C = \frac{1}{n}\Bigg[ \frac{\cos(p\pi/n^\prime) - \cos((p+1)\pi/n^\prime)
+ \cos(\pi/n^\prime)-1}{1-\cos(\pi/n^\prime)} \Bigg].  \label{cool}
\end{equation}
Recall that $n^\prime = n-p+1$.  Eq. (\ref{cool}) gives the largest value of $C$ consistent with our conditions, for fixed values of $n$ and $p$.  Note, for example, that when $n=5$ and $p=2$, 
Eq. (\ref{cool}) gives $C=\sqrt{2}/5$, just as we found before for this case.

We still need to optimize over $n$ and $p$.  It is best to make the block size $n$ very large---any state $w$ that is possible with block size $n$ is also allowed by block size $2n$---so we take the limit as $n$ goes to infinity.  Let $\alpha$ be the density of occupied sites---that is, $\alpha = p/n$---and let $n$ approach infinity with $\alpha$ held fixed.  In this limit, the concurrence becomes  
\begin{equation}
C_{\rm lim} = \frac{2}{\pi}(1-\alpha)\sin \Big( \frac{\alpha \pi}{1-\alpha} \Big).
\end{equation}
Taking the derivative, one finds that $C_{\rm lim}$ is maximized when 
\begin{equation}
\tan \Big( \frac{\alpha \pi}{1-\alpha} \Big) = \frac{\pi}{1-\alpha},
\end{equation}
which happens at $\alpha = 0.300844$, where $C_{\rm lim} = 0.434467$.
This is the highest value of concurrence that is consistent with our method of constructing the state of the chain and with our three conditions on $|\xi\rangle$.  Note that it is considerably larger than what we got in our first example, in which a string of singlets was mixed with a shifted version of the same string---one might call this earlier construction the ``bicycle chain'' state.  Unlike the bicycle chain state, our best state breaks the symmetry between the basis states $|0\rangle$ and $|1\rangle$: the fraction of qubits in the state $|1\rangle$ is about 30\% rather than 50\%.  Of course the entanglement would be just as large if the roles of $|1\rangle$ and $|0\rangle$ were reversed.  

It is interesting to ask what value of entanglement of formation the above value of concurrence corresponds to.  As a function of the concurrence, the entanglement of formation is given by
\begin{equation}
E_f = h \bigg( \frac{1+\sqrt{1-C^2}}{2} \bigg),
\end{equation}
where $h$ is the binary entropy function $h(x)=-[x\log_2 x +(1-x)\log_2 (1-x)]$.
For the above value of concurrence, one finds that the entanglement of formation is $E_f = 0.284934$ ebits.  (For the bicycle chain state, the entanglement of formation between neighboring pairs is only 0.118 ebits.)

If one can prove that this value is optimal, then it can serve as a reference point for interpreting entanglement values obtained for real physical systems.  A string of spin-1/2 particles interacting via the antiferromagnetic Heisenberg interaction, for example, has eigenstates that typically have some non-zero nearest-neighbor entanglement.  It would be interesting to find out how the entanglements appearing in these states compare to the maximum possible entanglement for a string 
of qubits.\footnote{Since the original version of
this paper was written, the question 
about the antiferromagnetic Heisenberg chain has been answered
for the ground state \cite{Oconnor}:
though the nearest-neighbor concurrence of the ground state 
is high ($C = 0.386$), it is not optimal.}

Clearly the problem we have analyzed here can be generalized.  One can consider a two or three-dimensional lattice of qubits and ask how entangled the neighboring qubits can be.  If we were to analyze these cases using assumptions similar to those we have made in the one-dimensional case, we would again find the problem reducing to a many-body problem, but with less tractable interactions.  Assuming that pairwise entanglement tends to diminish as the total entanglement is shared among more particles, one expects the optimal values of $C$ and $E_f$ to shrink as the dimension of the lattice increases.  

I would like to thank Kevin O'Connor for many valuable discussions on distributed entanglement.

\end{document}